%% file: paper.tex
\documentclass[
12pt,
a4paper,
amsmath,
superscriptaddress,
floatfix,
notitlepage,
longbibliography
]{revtex4-2}

\usepackage{graphicx,amssymb,amsmath,amsfonts}
\usepackage{epsfig}
\usepackage{xcolor}
\usepackage[mathscr]{eucal}
\usepackage{comment}
\usepackage{bm}
\usepackage{siunitx}
\usepackage{natmove}

\usepackage[utf8]{inputenc}
\usepackage{xr}
\let \ShowFixme = 0
\let \Date = 1
\include{defines}

\graphicspath{{./figures/}}

\begin{document}

\title{\papertitle}

\author{Taras Verkholyak}
\affiliation{Institute for Condensed Matter Physics, NASU, Lviv, Ukraine}

\author{Andrij Kuzmak}
\affiliation{Department for Theoretical Physics, I. Franko National University of Lviv, Ukraine}

\author{Svyatoslav Kondrat}
\affiliation{Institute of Physical Chemistry, Polish Academy of Sciences, 01-224 Warsaw, Poland}
\affiliation{Max-Planck-Institut f{\"u}r Intelligente Systeme, Heisenbergstra{\ss}e~3, D-70569 Stuttgart, Germany}
\affiliation{IV. Institut f\"ur Theoretische Physik, Universit{\"a}t Stuttgart,  Pfaffenwaldring 57, D-70569 Stuttgart, Germany}


\begin{abstract}

    Understanding charge storage in low-dimensional electrodes is crucial for developing novel ecologically friendly devices for capacitive energy storage and conversion, water desalination, \etc Exactly-solvable models allow in-depth analyses and essential physical insights into the charging mechanisms. So far, however, such analytical approaches have been mainly limited to lattice models. Herein, we develop a versatile, exactly-solvable, one-dimensional off-lattice model for charging single-file pores. Unlike the lattice model, this model shows an excellent quantitative agreement with three-dimensional Monte Carlo simulations. With analytical calculations and simulations, we show that the differential capacitance can be bell-shaped (one peak), camel-shaped (two peaks), or have four peaks. Transformations between these capacitance shapes can be induced by changing pore ionophilicity, cation-anion size asymmetry, or by adding solvent. We find that the camel-shaped capacitance, characteristic of dilute electrolytes, appears for strongly ionophilic pores with high ion densities, which we relate to charging mechanisms specific to narrow pores. We also derive a large-voltage asymptotic expression for the capacitance, showing that the capacitance decays to zero as the inverse square of the voltage, $C\sim u^{-2}$. This dependence follows from hard-core interactions and is not captured by the lattice model. 
	
\end{abstract}

\maketitle

\section{Introduction}

Confined ionic liquids attract a growing interest of the research community \cite{vatamanu:acsnano:15, Salanne2016Efficient, Li2018, Xu2020Computational, Simon2020Perspectives, Cruz2021Phase}, as they exhibit interesting physics and find numerous applications, particularly in green energy storage \cite{miller08a, simon08a, beguin14a, gonzalez16a} and conversion \cite{brogioli09a, hartel2015heat, Janssen2017Reversible}, water desalination \cite{porada13a, suss18a, zhang20a}, \etc For instance, the highest achievable capacitance has been obtained for supercapacitors with electrodes featuring subnanometer pores, comparable in size to the size of a desolvated ion \cite{gogotsi:sci:06, pinero:carbon:06, gogotsi:08}. Such an anomalous increase of capacitance is allegedly due to a superionic state emerging in narrow conducting pores, which screen the inter-ionic interactions \cite{rochester:cpc:13, Goduljan2014a, Mohammadzadeh2015a, Mohammadzadeh2016Interactions} and allow for tighter and more efficient \cite{merlet:natcom:13} packing of confined ions, thereby enhancing the capacitance \cite{kondrat:jpcm:11, Futamura2017}.

\label{ref2:p1:a}
Analytically tractable models are precious assets in physics, allowing one to unveil generic features and develop new physical insights more easily. The simplest exactly-solvable model for supercapacitor charging has been proposed by \citeauthor{kornyshev:fd:14} \cite{kornyshev:fd:14}, who mapped the charge storage in singe-file pores onto a one-dimensional (1D) Ising's spin model defined on a lattice. In this model, spins $+1$ and $-1$ mimic cations and anions and external magnetic field corresponds to the potential difference applied to a pore with respect to bulk electrolyte \cite{kornyshev:fd:14}. The assumption of nearest-neighbour interactions, essential to obtain the exact solution, seems reasonable due to the exponential screening of inter-ionic interactions in the superionic state. This 1D model describes the charging of electrodes based on carbon nanotubes (CNT) or CNT forest \cite{Pan2010Carbon, Muralidharan2018Molecular, Cao2019Highly} and has been developed to account for the presence of solvent/voids \cite{lee:prl:14}, ion-size asymmetry \cite{rochester:jpcc:16}, and multi-file pores \cite{ZaboronskyKornyshev2020Ising}.

Similar models have been applied in two dimensions to describe the charging of ultranarrow slit-shaped pores \cite{dudka:jpcm:16, Dudka2019Superionic, Groda2021Superionic}. While such pores are kinetically advantageous over single-file pores, the exact solutions of two-dimensional models are much more challenging to obtain and approximations are necessary \cite{dudka:jpcm:16, Dudka2019Superionic, Groda2021Superionic, kondrat:jpcm:11, lee:prx:16}. Due to the possibility of analytical solutions and lower computational costs of simulations, therefore, one-dimensional models have been a convenient testbed for gaining new physical insights into the supercapacitor charging \cite{kornyshev:fd:14, lee:nanotech:14, lee:prl:14, rochester:jpcc:16, Schmickler2015simple, pak_hwang:jpcc:16:IonTrapping, Schmickler2017capacitance, Schmickler2017Charge, qiao2018modeling, Kondrat2019a}. With some exceptions \cite{Schmickler2015simple, Frydel2019}, however, 1D analytical approaches have been limited to lattice models \cite{kornyshev:fd:14, lee:prl:14,rochester:jpcc:16, Horgan1DIsingCaoacitor:2012, Horgan1DIsingOverscreening:2012, Demery2016, Frydel2018, Kondrat2019a}. But how realistic is the lattice assumption? Do such models capture the essential physics of charge storage?

\label{ref2:p1:b}Herein, we propose a 1D off-lattice model for charging single-file pores, which has a generic solution applicable to various ionic systems. Motivated by 1D lattice models of \myrefs{kornyshev:fd:14, lee:prl:14, rochester:jpcc:16}, we map the charging of ultranarrow cylindrical pores onto a 1D multi-component system of hard rods with nearest-neighbour interactions, which has an exact solution \cite{longuet1958, Heying2004, Santos2007, Ben2009, santos2016:ch5}. In our work, we perform this mapping analytically and derive relations between the chemical potentials of 3D single-file and 1D lattice and off-lattice models, which allows a direct comparison among all models. To test them, we perform grand-canonical Monte Carlo simulations of 3D systems. With analytical calculations and simulations, we study the charging of single-file pores with equally-sized and size-asymmetric ions and non-polar solvents. We reveal that charging mechanisms, capacitance and stored energy are sensitive functions of pore ionophilicity, ion-size asymmetry and solvent concentration. Comparing these results with the results of the corresponding lattice model suggests that, in most cases, the lattice model does not capture the behaviour of the system qualitatively. Using the off-lattice model, we derive an asymptotic expression for the capacitance at large voltages, which is challenging to obtain from simulations or experiments. We discuss how this asymptotic behaviour relates to the asymptotic behaviours in other geometries.

\section{Theory and simulations}

\subsection{Model}

\begin{figure}[t]
\begin{center}
    \includegraphics[width=0.95\textwidth]{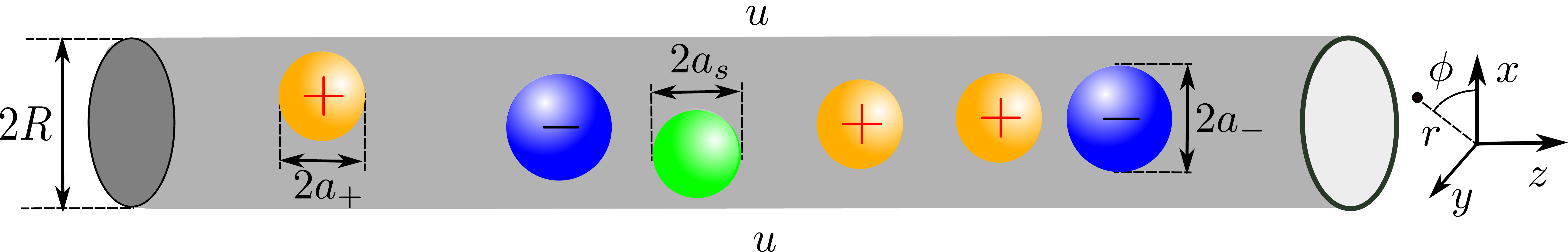}
	\caption{\ftitle{Model of a conducting cylindrical nanopore.} Ions and solvent molecules (if present) are confined to a narrow nanotube of radius $\porer$. Electrostatic potential difference $u$ is applied to the nanotube with respect to bulk electrolyte (not shown). The ion and solvent radii are $\ionr_\pm < \porer$ and $\solvr < \porer$, respectively. In 1D models (\sects{sec:model:offlat}{sec:model:lat}), the centers of all molecules are located on the symmetry axis of the nanotube. In 3D Monte Carlo simulations, the ions and solvent can take any position inside the pore provided it is not sterically prohibited. Polar coordinates $r$ and $\phi$ are defined  within the $(x,y)$ plane perpendicular to the tube symmetry axis. 
    \label{fig:model}
}
\end{center}
\end{figure}

We consider an ionic liquid and solvent (if present) confined to a cylindrical nanopore of a supercapacitor electrode (\fig{fig:model}). We model ions and solvent as charged and neutral hard spheres, respectively. An electrostatic interaction energy between ions of types $\alpha$ and $\gamma$ located at $\rr_1=(0, r_1, 0)$ and $\rr_2=(z, r_2, \varphi)$ (in cylindrical coordinates, \fig{fig:model}) inside a pore of radius $\porer$ is \cite{rochester:cpc:13}
\begin{multline}
    \label{eq:Uel:full}
	\beta \psi_{\alpha\gamma} (\rr_1, \rr_2) 
	= \beta \psi_{\alpha\gamma} (z, r_1, r_2, \phi) \\
	=  \frac{2 \lB q_\alpha q_\gamma}{\porer} \sum_{m=0}^\infty A_n \cos(m\phi)
        \sum_{n=1}^\infty \frac{J_m (k_{nm} r_1/ \porer) J_m (k_{nm} r_2/ \porer)}{k_{nm}[Y_{m+1}(k_{nm})]^2} e^{-k_{nm} z/\porer},
\end{multline}
where $q_\alpha$ and $q_\gamma$ are the ion charges (in units of the proton charge $e$), $A_0=1$ and $A_{m} = 2$ for $m\ne 0$, $J_m$ and $Y_m$ are the Bessel functions of the first and second kind and $k_{nm}$ is the $n$th positive root of $J_m$. The Bjerrum length $\lB = \beta e^2/ \varepsilon$ (in Gaussian units), where $\varepsilon$ is the dielectric constant inside the pore and $\beta=(k_B T)^{-1}$ the inverse temperature ($k_B$ is the Boltzmann constant and $T$ temperature). For two ions on the symmetry axis of the nanotube, $r_1=r_2 = 0$, \eq{eq:Uel:full} simplifies to
\begin{align}
    \label{eq:Uel:axis}
	\beta \psi_{\alpha\gamma} (z) =  \frac{2\lB q_\alpha q_\gamma}{\porer} 
        \sum_{n=1}^\infty \frac{e^{-k_{n0} z/\porer}}{k_{n0}[J_{1}(k_{n0})]^2},
\end{align}
which for large ion-ion separations, $z \gg \porer$, becomes
\begin{align}
    \label{eq:Uel:approx}
	\beta \psi_{\alpha\gamma} (z) \approx  \frac{3.08\lB q_\alpha q_\gamma}{\porer} \e^{-2.4 z/\porer}.
\end{align}
\Eq{eq:Uel:approx} approximates \eq{eq:Uel:axis} in a wide range of parameters remarkably well \cite{kornyshev:fd:14}. It demonstrates an exponential screening of electrostatic interactions in metallic nanotubes and has motivated several exactly-solvable 1D lattice models with nearest neighbour interactions \cite{kornyshev:fd:14, lee:prl:14, rochester:jpcc:16, ZaboronskyKornyshev2020Ising} (\sect{sec:model:lat}).

\label{ref1:eq3}Although \eqss{eq:Uel:full}, \eqref{eq:Uel:axis}, and \eqref{eq:Uel:approx} have been derived for a perfectly metallic cylinder, recent quantum density functional (DFT) calculations \cite{Goduljan2014a, Mohammadzadeh2015a, Mohammadzadeh2016Interactions} suggest that they provide a reliable approximation of inter-ionic interactions also inside gold and carbon nanotubes. It is interesting to note that the effective value of $\porer$ fitting \eq{eq:Uel:approx} to the DFT calculations was even smaller than the physical pore radius, implying stronger screening.

An interaction energy between ion $\alpha$ and the pore wall due to image charges is \cite{rochester:cpc:13}
\begin{multline}
	\label{eq:Uself}
	\beta \psi_\alpha (\rr) = \beta \psi_\alpha (r) \\
	= \frac{\lB q_\alpha^2}{2 \pi \porer} \sum_{m=0}^\infty a_m \int_0^{2\pi} d \phi \cos(m\phi)
	\int_0^\infty d\xi \frac{I_m (\xi r/\porer)}{I_m(\xi)} K_0 \left(\frac{\xi}{r}\sqrt{r^2 + \porer^2-2r\porer \cos\phi}\right),
\end{multline}
where $r$ is the radial coordinate (\fig{fig:model}), and $I_m(x)$ and $K_m(x)$ are the modified Bessel functions of the first and second kind, respectively.

\subsection{Partition function and Monte Carlo simulations}

Thermodynamic properties of a $M$-component system can be obtained from the grand partition function
\begin{align}
	\label{eq:Xi}
    \Xi = \sum_{N_1=0}^\infty \cdots \sum_{N_M=0}^\infty \e^{\beta \sum_{\alpha=1}^M\mu_\alpha N_\alpha} Z(\{N_\alpha\}),
\end{align}
where $\mu_\alpha$ is the electrochemical potential of particles of type $\alpha$ (note that we have included the thermal de Broglie wavelength into $\mu_\alpha$), and 
\begin{align}
	\label{eq:Z}
	Z (\{N_\alpha\}) = \prod_{\alpha=1}^M \frac{1}{N_\alpha!}
		\prod_{i_\alpha=1}^{N_\alpha} \int  d^3r_{i_\alpha} \e^{-\beta\Psi}
\end{align}
is the configurational partition function of a canonical ensemble of $\{N_\alpha\}$ particles,
\begin{align}
	\Psi= \Psi_1 + \Psi_2 
	= \sum_\alpha \sum_{i_\alpha}\psi_{\alpha}(\rr_{i_\alpha})
	+\frac{1}{2}\sum_{\alpha\gamma}\sum_{i_\alpha j_\gamma}{\vphantom{\sum}}^\prime \psi_{\alpha\gamma} (\rr_{i_\alpha}, \rr_{j_\gamma}) 
\end{align}
is the total potential energy of the system and $\sum^\prime$ means that self-interactions are excluded, implying $i_\alpha\ne j_\alpha$. In this article, we consider $M=2$ to describe charging with neat ionic liquids (\sects{sec:res:IL}{sec:res:ILasym}) and $M=3$ for a mixture of ions and solvent (\sect{sec:res:ILsolv}).

Calculating grand partition function \eqref{eq:Xi} analytically is a formidable task. We have performed grand canonical Monte Carlo (MC) simulations to compute the thermodynamic characteristics of our model numerically. In this model, we used potentials \eqref{eq:Uel:full} to describe interactions between ions and potentials \eqref{eq:Uself} for the ions and the nanotube surface. In addition, all particles (ions and solvent if any) interacted sterically with each other and with the nanotube. Our simulations consisted of translational, Widom insertion/deletion \cite{widomMC:1963} and molecular-type swap \cite{kondrat:pccp:11} moves. We performed $10^6$ MC steps for equilibration and, depending on a system, from $10^6$ to $10^7$ steps in production runs. Periodic boundary conditions were applied in the $z$ direction. In all simulations, the tube length was $L=\SI{25}{\nano\meter}$.

\subsection{Reduction to 1D off-lattice model}
\label{sec:model:offlat}

We consider a nanotube so narrow that it can admit only a single row of ions. To obtain a 1D model, we neglect the radial dependence of the ion-ion interactions (\eq{eq:Uel:full}), assuming that the interaction potential depends only on the axial distance between the ions, as given by  \eq{eq:Uel:axis}. However, we retain the radial dependence of the self-energy, \eq{eq:Uself}. In this case, partition function \eqref{eq:Z} reduces to
\begin{multline}
	\label{eq:Z:1D}
	Z (\{N_\alpha\}) = \prod_{\alpha=1}^M\frac{1}{N_\alpha!}
		 \prod_{i_\alpha=1}^{N_\alpha} \int d^2 r_{i_\alpha}\e^{-\beta \psi_{\alpha}(r_{i_\alpha})}
		 \prod_{i_\alpha=1}^{N_\alpha} \int_0^L d z_{i_\alpha} \e^{-\beta \Psi_2(\{z_{i_\alpha}\})}\\
	= \prod_{\alpha=1}^M \e^{-\beta \bar \psi_\alpha N_\alpha} Z_{\oneD} (\{N_\alpha\}),
\end{multline}
where $L$ is the pore length, $\int d^2r_{i_\alpha} = \int_0^{2\pi} d \phi_{i_\alpha} \int_0^{\porer-2\ionr_\alpha} r_{i_\alpha} d r_{i_\alpha}$ denotes integration over the area perpendicular to the nanotube symmetry axis and accessible to particle $i$ of type $\alpha$,
\begin{align}
	\label{eq:psi_bar}
	\beta \bar \psi_\alpha = \ln\left(2 \pi \int_0^{\porer-\ionr_\alpha} \e^{-\beta \psi_\alpha(r)} r dr \right),
\end{align}
and
\begin{align}
	\label{eq:Z:1D:final}
	Z_{\oneD} (\{N_\alpha\}) = \prod_{\alpha=1}^M  \frac{1}{N_\alpha!}
	\prod_{i_\alpha=1}^{N_\alpha} \int_0^L d z_{i_\alpha} \e^{-\beta \Psi_2(\{z_{i_\alpha}\})}
\end{align}
is the partition function of a 1D system of $\{N_\alpha\}$ particles, assuming that all particles reside on the symmetry axis of the nanotube and hence $\Psi_2$ depends only on the axial distances $z_{i_\alpha}$. This gives for the grand partition function of an effective 1D model
\begin{align}
	\label{eq:Xi:1D}
	\Xi_{\oneD} = \sum_{N_1=0}^\infty \cdots \sum_{N_M=0}^\infty \e^{\beta \sum_\alpha \mu_\alpha^{(\oneD)} N_\alpha} Z_{\oneD}(\{N_\alpha\}),
\end{align}
where
\begin{align}
\mu_\alpha^{(\oneD)} = \mu_\alpha - \bar \psi_\alpha
\end{align}
is the chemical potential of particle type $\alpha$ in the 1D model. Note that in the absence of particle-wall interactions (\ie, for $\psi_\alpha = 0$), $\bar \psi_\alpha = \ln \left[\pi (\porer-\ionr_\alpha)^2\right]$ does not vanish and reduces to the cross-sectional area accessible to particles of type $\alpha$.

\subsection{Exact solution of the 1D off-lattice model}

The 1D partition function, \eq{eq:Xi:1D}, is still challenging to calculate analytically. However, an exact analytical solution exists if we assume that only neighbouring  particles interact with each other \cite{longuet1958}. This is a reasonable assumption due to the strong exponential screening of inter-ionic interactions inside ultranarrow metallic nanotubes (\eq{eq:Uel:approx}). 

The exact solutions for classical 1D systems of particles have a long history, starting from the seminal work by \citeauthor{Tonks1936Complete} \cite{Tonks1936Complete}, who considered the statistical mechanics of one-dimensional hard-rods. The Tonks model was generalized by \citeauthor{Takahashi1942} \cite{Takahashi1942}, who additionally included short-ranged interactions, restricted to neighboring particles, which is possible to do in 1D. These solutions have been further generalized to binary \cite{Kikuchi1955} and multicomponent \cite{longuet1958, Heying2004, Santos2007, Ben2009, santos2016:ch5} systems. Here, we present a derivation for an $M$-component system in the grand canonical ensemble, based on \myref{longuet1958}.

To calculate the thermodynamic quantities assuming the nearest-neighbours interactions, we first take the Laplace transform of $\Xi_\oneD$
\begin{align}
	\label{eq:XiL}
	\XiL (\{\mu_\alpha\}, T, s) = \int_0^\infty dL \e^{-sL} \Xi_\oneD (\{\mu_\alpha\}, T, L).
\end{align}
Since the grand thermodynamic potential $J(\{\mu_\alpha\}, T, L) = p L = - k_B T \ln \Xi_\oneD$, where $p$ is pressure and $L$ the nanotube length, it follows from \eq{eq:XiL} that $\XiL$ diverges for $s \le \beta p$. We will use this condition later to derive the equation of state. 

The computation of $\XiL$ is tedious and the details are presented in \ssect{sec:derivation}. The main steps involve rewriting the partition function as a sum over the total number of particles and making use of the nearest-neighbour assumption, which allows one to calculate $\XiL$ in terms of the `transfer matrix' $\Theta$
\begin{align}
	\label{eq:XiL2}
	\XiL(\{\mu_\alpha\}, T, s) = \frac{1}{s} + \frac{1}{s^2} \sum_\alpha \e^{\beta \mu_\alpha}
	- \sum_{N>1} \frac{1}{N} \frac{\partial}{\partial s}\Tr \left[\Theta(\{\mu_\alpha\}, T, s)\right]^N,
\end{align}
where the first and second terms are due to the empty pore and a single in-pore ion, respectively, and the matrix elements are
\begin{align}
	\Theta_{\alpha\gamma}(\{\mu_\alpha\}, T, s) = \left( \e^{\beta \mu_\alpha} \e^{\beta \mu_\gamma}\right)^{1/2}
	\int_0^\infty dz \e^{-sz - \beta \psi_{\alpha\gamma}(z)}.
\end{align}
One can rewrite \eq{eq:XiL2} as follows 
\begin{align}
	\label{eq:XiL3}
	\XiL(\{\mu_\alpha\}, T, s) = \frac{1}{s} + \frac{1}{s^2}  \sum_\alpha \e^{\beta \mu_\alpha}
	-  \sum_\alpha \frac{\theta}{1-\theta_\alpha} \frac{\partial \theta_\alpha}{\partial s},
\end{align}
where $\theta_\alpha$ is the $\alpha$'s eigenvalue of $\Theta$ and we have summed over $N$ assuming that the sum is convergent. Since $\theta_\alpha = \theta_\alpha(\{\mu_\alpha\}, T, s)$ are analytical functions, this sum and hence $\XiL$ diverge only when at least one of $\theta_\alpha \to 1$. Thus, the value of $s$ at which $\theta_\alpha (\{\mu_\alpha\}, T, s) = 1$ corresponds to the equilibrium pressure, $p = k_B T s$, as mentioned below \eq{eq:XiL}. The condition $\theta_\alpha = 1$ can be obtained from the characteristic equation for the transfer matrix
\begin{align}
\label{eq:det}
	\Delta(\{\mu_\alpha\}, T, s)=
\left|
\begin{array}{ccc}
\Theta_{11}-1 & \Theta_{12}   & \dots\\
\Theta_{12}   & \Theta_{22}-1 & \dots\\
\dots & \dots & \dots 
\end{array}
\right|=0. 
\end{align}
The solution of this equation, $s = f(\{\mu_\alpha\}, T)$, therefore, gives the equation of state $p = k_B T f(\{\mu_\alpha\}, T)$. 

To obtain particle densities, we take the full differential of \eq{eq:det}, using $s=p/k_BT$,
\begin{align}
	\sum_{\alpha}\frac{\partial\Delta}{\partial\mu_\alpha}d\mu_\alpha + \frac{\partial\Delta}{\partial T}dT + \frac{\partial\Delta}{\partial p}dp=0.
\end{align}
Comparing this equation to the Gibbs-Duhem equation,
\begin{align}
	\sum_{\alpha} \bar N_\alpha d\mu_\alpha + SdT -Ldp=0,
\end{align}
where $S$ is entropy and $\bar N_\alpha$ the average number of particles of type $\alpha$, we obtain the following relations
\begin{align}
	\bar N_\alpha:S:L= \frac{\partial\Delta}{\partial\mu_\alpha}:\frac{\partial\Delta}{\partial T}:-\frac{\partial\Delta}{\partial p}. 
\end{align}
The particle density is, therefore,
\begin{align}
\label{eq:densities}
	\rho_\alpha= \frac{\bar N_\alpha}{L} = 
	- \frac{\partial\Delta/\partial\mu_\alpha}{\partial\Delta/\partial p}
	=- k_BT \frac{\partial\Delta/\partial\mu_\alpha}{\partial\Delta/\partial s}.
\end{align}

\subsubsection{Application to ionic liquids}

We now apply this solution to ionic liquids in a single-file pore. For simplicity of presentation, we consider monovalent ions of the same size  and write for their electrochemical potential
\begin{align}
	\mu_\pm = \muIL \pm e u,
	\label{eq:mupm}
\end{align}
where $\muIL$ is the chemical potential of ions (assumed the same for cations and anions) and $u$ is the voltage applied to the electrode with respect to bulk electrolyte. \Eq{eq:det} gives
\begin{align}
	\label{eq:det:IL}
	\e^{2\beta \muIL} (\eta_{++}^2 - \eta_{+-}^2) - 2 \e^{\beta \muIL} \cosh (\beta e u) + 1 = 0,
\end{align}
where
\begin{align}
	\eta_{\alpha\gamma}(s) = \int_0^\infty dz \e^{-sz - \beta \psi_{\alpha\gamma}(z)}
\end{align}
and $\psi_{\alpha\gamma}$ is given by \eq{eq:Uel:axis} or \eqref{eq:Uel:approx} (in our calculations, presented below, we have used \eq{eq:Uel:approx}). Note that $\eta_{+-}=\eta_{-+}$ and that $\eta_{++}=\eta_{--}$ for symmetric systems considered here. One can solve \eq{eq:det:IL} for $u$ obtaining for positive/negative voltages
\begin{align}
	\label{eq:u:s}
	u(s)=\frac{1}{\beta e}{\rm acosh}
	\left(
	\frac{\e^{\beta \muIL}(\eta_{++}^2-\eta_{+-}^2) \pm \e^{-\beta \muIL}}{2\eta_{++}}.
	\right)
\end{align}
\Eq{eq:densities} gives for ion densities
\begin{align}
	\label{eq:rho:s}
	\rho_\pm (s)
	=-\frac{1}{2} \frac{\e^{\beta \muIL}(\eta_{++}^2-\eta_{+-}^2) -\e^{\pm \beta e u} \eta_{++}}
	{\e^{\beta \muIL}(\eta_{++}\eta'_{++}-\eta_{+-}\eta'_{+-}) - \cosh(\beta e u)\eta'_{++}},
\end{align}
where $\eta'_{\alpha\gamma} = \partial \eta_{\alpha\gamma}/\partial s$. \Eqs{eq:u:s}{eq:rho:s} are parametric equations for the in-pore ion densities as functions of the applied potential difference $u$. The solution for an IL-solvent mixture is more lengthy and is presented in \ssect{sec:offlat:ILsolv}.

\subsection{Reduction to 1D lattice model}
\label{sec:model:lat}

To map our system onto a lattice model, we assume $L=2 \ionr \Ncells $ and divide the integration region in \eq{eq:Z:1D} into $\Ncells$ cells of size $2\ionr$, where $\ionr=\ionr_\pm=\solvr$ is the radius of ions and solvent molecules, taken here the same (for a lattice model with asymmetric ion/solvent sizes, see \myref{rochester:jpcc:16}). We restrict our considerations to neat ILs and ion-solvent mixtures so that a cell can be either empty or occupied by an ion or solvent if present. At each lattice site, we introduce a state variable $\sigma_i=\{0, \pm, \solv\}$, where $\sigma_i= 0$ denotes an empty site and $\sigma_i = \pm $ ($\sigma_i=\solv$) means that site $i$ is occupied by a $\pm$ ion (solvent). To calculate the inter-ionic interactions, we assume that the separation between two ions from the neighbouring cells is equal to the lattice constant $2\ionr$. We then obtain for the grand partition function
\begin{align}
	\label{eq:Xi:lat}
	\Xi_\lat = \prod_i \sum_{\sigma_{i=\{0,\pm,\solv\}}} 
		\exp\left\{- \beta \sum_{i < j} \psi_{\sigma_i\sigma_j} (2\ionr|i-j|) + \beta \sum_i \mu_{\sigma_i}^{(\lat)} \right\},
\end{align}
where the product runs over all lattice sites (\ie, $i=1,2,\cdots,\Ncells$), $\psi_{\sigma_i\sigma_j} (z)$ is given by \eq{eq:Uel:axis} when $\sigma_i,\sigma_j = \{\pm\}$ and is zero otherwise, and 
\begin{align}
	\label{eq:mulat}
	\mu_{\sigma_i}^{(\lat)} = \mu^{(\oneD)}_{\sigma_i} - k_BT \ln (2\ionr)
\end{align}
except for $\mu^{(\lat)}_0 = 0$. Since solvent does not interact with other particles other than via excluded volumes, we can sum up over solvent/void. Introducing `spin' variable $S_i = \{0, \pm 1\}$ and limiting the interactions to nearest neighbours, we arrive at
\begin{align}
	\label{eq:Xi:lat2}
	\Xi_\lat = \prod_i \sum_{S_i=\{0, \pm 1\}} \e^{- \beta H(\{S_i\})}
\end{align}
where  
\begin{align}
	\label{eq:H}
	H = J \sum_i S_i S_{i+1} 
	- \mu_+^{(\lat)} \sum_i \frac{S_i^2 + S_i}{2}
	- \mu_-^{(\lat)} \sum_i \frac{S_i^2 - S_i}{2}
	- \mu_0^{(\lat)} \sum_i (1-S_i^2),
\end{align}
the coupling constant $J= \psi_{++}(2\ionr)>0$, and the renormalized chemical potential of zero component
\begin{align}
	\label{eq:mulat:0}
	\mu_{0}^{(\lat)} = k_BT \ln \left( 1 + \e^{\beta \mus^{(\lat)}}\right).
\end{align}
Hamiltonian \eqref{eq:H} is the Blume-Capel model, well-known in the theory of magnetism, and has an exact analytical solution in 1D \cite{Blume66, Capel66}. This model is equivalent to the model of \myref{lee:prl:14} developed for charging single-file pores, except that we have introduced explicitly the chemical potential of solvent ($\mus^{(\lat)}$), which allows us to distinguish between solvent and void. In particular, the solvent and ion densities are
\begin{align}
	\rhos = -\frac{1}{\beta L}\frac{d \ln \Xi_\lat}{d \mus^{(\lat)}} 
	\qquad \textrm{and} \qquad 
	\rho_\pm = - \frac{1}{\beta L}\frac{d \ln \Xi_\lat}{d \mu_\pm^{(\lat)}},
\end{align}
while the density of voids is $\rhomax -\rhos-\rho_+-\rho_-$, where $\rhomax= (2a)^{-1}$ is the maximum 1D density (\cf \figs{fig:solv:nonpol}{fig:solv:charging}). Exact results for this model are summarised in \ssect{sec:sollattmodel}.

\section{Results and discussion}

We have used the exact solutions of the 1D lattice and off-lattice models to study the charging of single-file pores and tested these analytical results with 3D Monte Carlo (MC) simulations. In all calculations and simulations, temperature was $T=\SI{300}{\kelvin}$ and the in-pore relative dielectric constant $\varepsilon = 2.5$, giving the Bjerrum length $\lB = \SI{22.2}{\nano\meter}$. In most cases, we considered ions and solvent molecules of the same radius $\ionr = \SI{0.25}{\nano\meter}$, except for \sect{sec:res:ILasym}, where we discuss ion-size asymmetry.

\subsection{Neat ionic liquids}
\label{sec:res:IL}

\subsubsection{Uncharged pores}

\begin{figure}[t]
\begin{center}
    \includegraphics[width=0.8\textwidth]{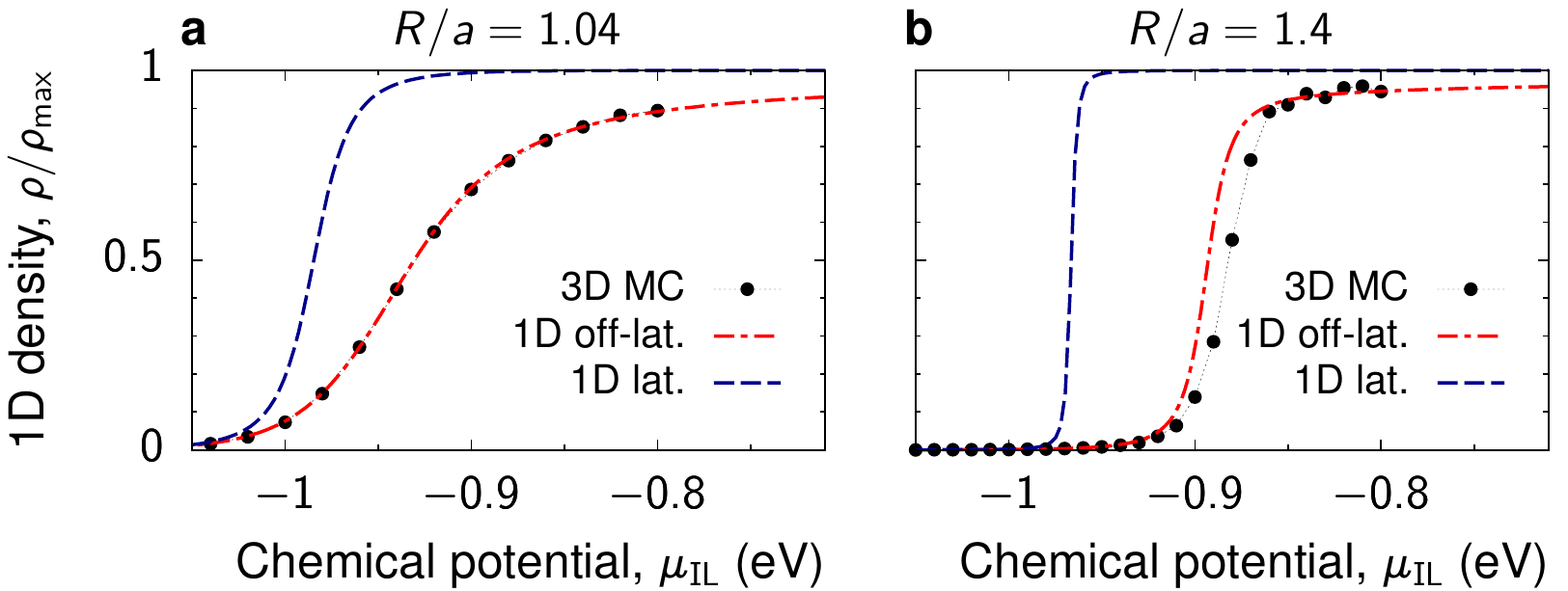}
	\caption{\ftitle{Ionic liquids in uncharged metallic nanotubes.} 1D ion density is shown as a function of ion chemical potential $\muIL$ for the pore radius $R=\SI{0.26}{\nano\meter}$ ($R/a = 1.04$) in \fsub{a} and for $R=\SI{0.35}{\nano\meter}$ ($R/a = 1.4$) in \fsub{b}. Ion radii $\ionr = \ionr_\pm = \SI{0.25}{\nano\meter}$, temperature $T=\SI{300}{\kelvin}$ and the in-pore dielectric constant $\varepsilon = 2.5$. The 1D off-lattice model and 3D MC simulations agree well for $R/a \lesssim 1.4$ (see also \sfig{fig:charging:pore_size}), while the 1D lattice model deviates from them qualitatively in all cases considered, except for strongly ionophobic pores.
    \label{fig:IL:nonpol}
}
\end{center}
\end{figure}

We first discuss monovalent ions in uncharged pores. In \fig{fig:IL:nonpol}, we show how the in-pore ion density varies with the chemical potential of ions $\muIL$ (\eq{eq:mupm}). The ion density increases with $\muIL$ more steeply when the pore gets wider, which is because the attraction between cations and anions becomes stronger, making it more favourable for ions to enter the pore. For ultranarrow pores ($\porer/\ionr = 1.04$), the off-lattice model shows an excellent agreement with the simulations, but it starts to deviate from the MC results as the pore width increases. Nevertheless, the agreement is good up to $\porer/\ionr = 1.4$ (\fig{fig:IL:nonpol}b) and the off-lattice model predicts qualitatively correct behaviour even for wider pores (\sfig{fig:charging:pore_size}).

In sharp contrast, the results of the 1D lattice model differ considerably from the 3D MC results, even for $\porer/\ionr\approx 1$. The ion density increases sharply with the chemical potential and quickly saturates at the maximum density $\rhomax= (2\ionr)^{-1}$, while in the off-lattice model/MC simulations, the ion density approaches $\rhomax$ much more slowly. \label{ref2:p2:a} This difference arises because the lattice model overestimates the interaction energy and underestimates the entropic contribution, which, in fact, diverges when $\rho \to \rhomax$ (\cf \eq{eq:s_1d_rods} below).

\subsubsection{Charging}

\begin{figure}[t]
\begin{center}
    \includegraphics[width=0.95\textwidth]{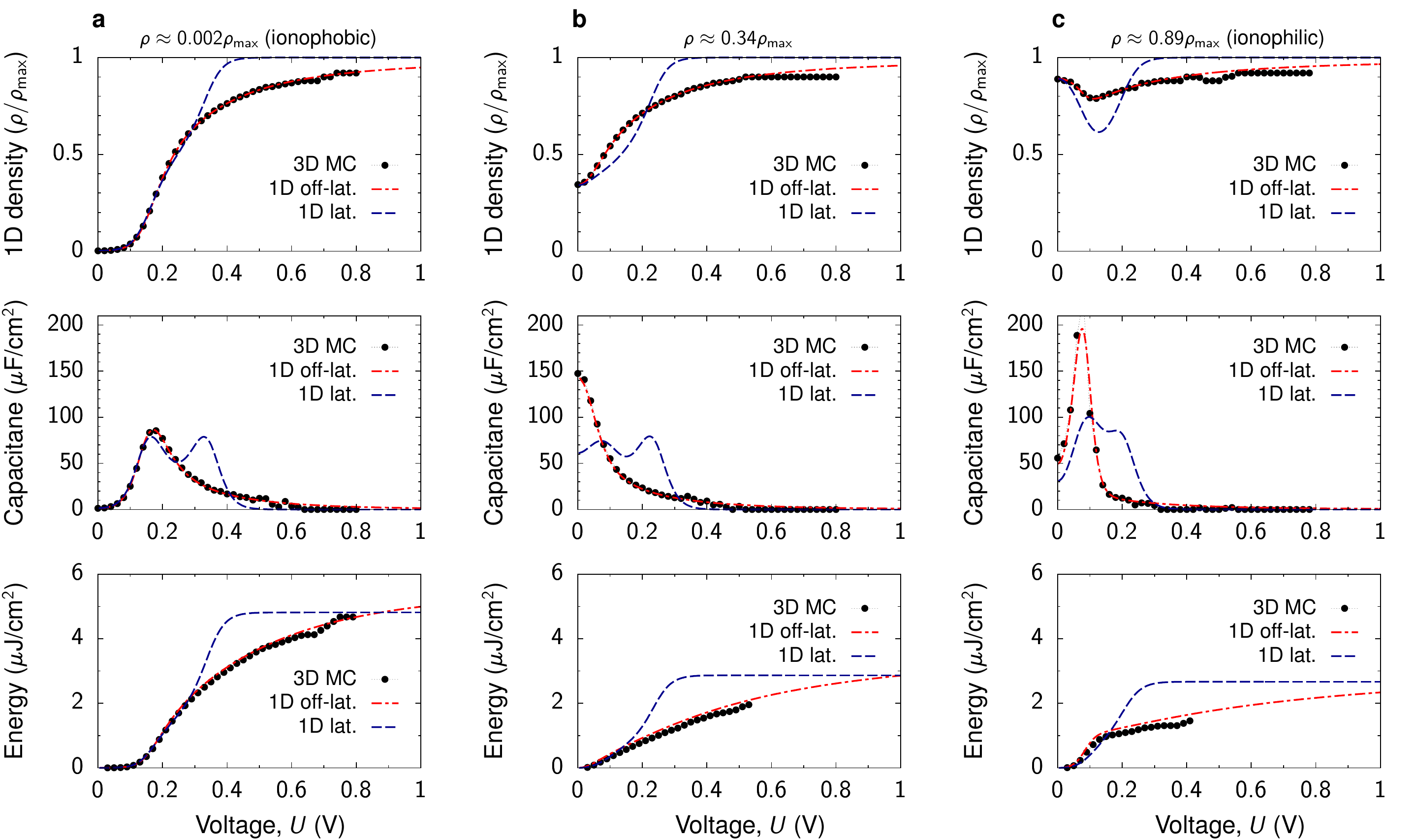}
	\caption{\ftitle{Charging metallic nanotubes with ionic liquids.} 1D ion density  (top row), differential capacitance (middle row) and stored energy density (bottom row) as functions of applied potential difference for \fsub{a} ionophobic pore with 1D in-pore ion density $\rho/\rhomax = 0.002$ ($\rhomax = (2\ionr)^{-1}$ and $\ionr = \SI{0.25}{\nano\meter}$ is the ion radius); \fsub{b} moderately filled pore with $\rho/\rhomax = 0.34$; \fsub{c} ionophilic pore with $\rho/\rhomax = 0.94$ at zero voltage. The chemical potentials are $\muIL=\SI{-1.1}{eV}$, $\SI{-0.95}{eV}$ and $\SI{-0.8}{eV}$, respectively. In the lattice model, the chemical potentials have been adjusted to give the same densities as in 3D MC simulations: $\muIL^{(\lat)}=\SI{-1.096}{eV}$, $\SI{-0.991}{eV}$ and $\SI{-0.961}{eV}$, respectively. The 1D off-lattice model shows an excellent agreement with the 3D MC results, while the 1D lattice model predicts a qualitative different behaviour in all cases except for the ionophobic pore at low voltages (panel (a)). Ion radii $\ionr = \ionr_\pm =\SI{0.25}{\nano\meter}$, pore radius $\porer=\SI{0.26}{\nano\meter}$ ($\porer/\ionr = 1.04$), temperature $T=\SI{300}{\kelvin}$, and the in-pore dielectric constant $\varepsilon=2.5$. See \sfigs{fig:IL:charging:R3.5}{fig:IL:charging:R4} for $\porer/\ionr = 1.4$ and $\porer/\ionr = 1.6$.
    \label{fig:IL:charging}
}
\end{center}
\end{figure}

We consider three values of $\muIL$ corresponding to three types of pores: ionophobic, which are nearly free of ions at zero voltage; weakly ionophobic ($=$ weakly ionophilic), which are moderately filled with ions; and ionophilic pores, which are nearly fully occupied by ions. To facilitate a comparison between different models, we chose the chemical potential in the lattice model to provide the same in-pore ion density at zero voltage as obtained with MC simulations. 

We have again found an excellent agreement between the 1D off-lattice model and 3D MC simulations. In contrast, the lattice model agrees with the MC results only for ionophobic pores at low voltages (\fig{fig:IL:charging}a). \label{ref2:p3:a}In this case, the counter-ion density is low, which means that the counter-ions practically do not interact with each other due to the strong screening of inter-ionic interactions (\eq{eq:Uel:approx}). Hence, both 1D lattice and off-lattice models approximate the nanotube charging quite well. \label{ref2:p2:b} As the voltage increases, however, the differences between the two models become apparent. These differences are due to the discrete nature of the lattice model and the entropy of ion packing (see \eq{eq:s_1d_rods} below), which the lattice model does not take into account. It has important consequences for large-voltage charging, which we discuss in \sect{sec:asymp}.

Unlike ionophobic and moderately-filled pores, the low-voltage charging of strongly ionophilic pores proceeds via (co-)ion desorption (\fig{fig:IL:charging}c). This effect is captured by the lattice model qualitatively, although there are quantitative differences.

An experimentally measurable quantity is the differential capacitance \cite{limmer:prl:13, fedorov_kornyshev:chemrev:14}
\begin{align}
	\label{eq:cap}
	C(u) = \frac{dQ}{du} = 
	\frac{\beta e^2}{2\pi \porer L} 
		\left[ \langle (N_+-N_-)^2\rangle - (\bar N_+ - \bar N_-)^2\right],
\end{align}
where $Q = e[\rho_+ - \rho_-]/(2\pi\porer)$ is the charge (per surface area) accumulated in a pore,  $\bar N_\pm = \langle N_\pm\rangle$ is the average number of $\pm$ ions, and $\langle\cdot\rangle$ denotes thermal averaging. \Fig{fig:IL:charging} (middle row) shows that the capacitance transforms from camel-shaped (minimum at zero voltage) to bell-shaped (maximum at zero voltage) and back to camel-shaped as the pore ionophilicity (in-pore ion density) increases. This behaviour is unlike flat electrodes, which only show a camel-to-bell transition for increasing ion density \cite{dicaprio:03, dicaprio:04, kornyshev:07}. The re-entrant bell-to-camel transformation in the nanopore is plausibly due to co-ion desorption, which does not occur at flat electrodes. This behaviour should be generic to ultranarrow pores and also emerge for narrow slit confinements. The lattice model does not reproduce this sequence of capacitance shapes. Instead, it predicts a transformation from two-peak capacitance to camel-shaped capacitance for the increasing pore ionophilicity (\sfig{fig:IL:charging:lat}). The second peak for ionophobic pores can be related to a particle-hole duality of the lattice model \cite{lee:prl:14}, which is absent in the continuous model.

\subsubsection{Large-voltage asymptotic behaviour}
\label{sec:asymp}

The analytical solutions allowed us to calculate asymptotic behaviours at large voltages, which is challenging to obtain from simulations or experiments. The lattice model predicts that the capacitance decays exponentially to zero as $\beta e u \to \infty$ (\ssect{asymplatt}). This exponential decay shows up in the off-lattice model as well. However, the off-lattice model brings also power-law and logarithmic dependences (\ssect{sec:offlatt:asymp})
\begin{align}
	\label{eq:C:asymp}
    C(u) \approx \frac{k_BT}{4\pi \porer \ionr} \left[ \frac{1}{u^2} 
	+ \frac{2k_BT}{e} \frac{\ln(e u/k_BT)}{u^3} \right]
	+ O\left(u^{-3}\right).
\end{align}
It is noteworthy that \eq{eq:C:asymp} follows from the expansion of the Lamber $W$-function $W(x)$ at large $x$, which converges slowly \cite{Krynytskyi2021}, implying that higher order terms might be important at intermediate and even large voltages (\sfig{fig:IL:charging:asymp}).

Notably, asymptotic expansion \eqref{eq:C:asymp} does not involve the parameters of electrostatic interactions, such as pore radius $\porer$ or dielectric constant ($\porer$ appears in \eq{eq:C:asymp} to normalise capacitance to surface area). This observation motivates a simple phenomenological model to rationalise the $1/u^{2}$ asymptotic behaviour. Assuming that only counter-ions are present in a nanotube, we write for the free energy density
\begin{align}
	\label{eq:phen_free_energy}
	f(\rhoc) = -e u \rhoc - T s (\rhoc),
\end{align}
where $\rhoc$ is the counter-ion density and $s(\rhoc)$ is the entropy density. Note that \eq{eq:phen_free_energy} does not contain electrostatic or any other interactions between the ions. We now use for $s(\rhoc)$ the exact solution for a 1D hard-rod fluid; up to a linear term in $\rhoc$, which merely shifts $u$, it is given by \cite{Tonks1936Complete}
\begin{align}
	\label{eq:s_1d_rods}
	s(\rhoc)  = - k_B \rhoc \ln (\rhoc^{-1}- \rhomax^{-1}),
\end{align}
where $\rhomax=(2\ionr)^{-1}$ is the maximum achievable density. It is noteworthy that $s(\rhoc)$ diverges logarithmically for $\rhoc \to \rhomax$. At high densities, $f \approx -eu\rhoc + k_BT \rhomax \ln (\rhomax-\rhoc)$ up to the leading order in $\rhomax-\rhoc$. Minimizing $f$ with respect to $\rhoc$, we find for the accumulated charge $e \rhoc \approx e \rhomax - \rhomax k_BT/u$, which gives the first term in \eq{eq:C:asymp} upon differentiating with respect to $u$ and normalizing to the surface area. Thus, the large-voltage asymptotic behaviour is determined solely by hard-rod entropic interactions. Such interactions are absent in the lattice model, explaining why this model does not predict the power-law decay of the capacitance.

\subsubsection{Energy storage}
\label{sec:E}

We also calculated the energy stored in a nanopore
\begin{align}
	\label{eq:E}
	E(U) = \int_0^U C(u) u du.
\end{align}
\Fig{fig:IL:charging} (bottom rows) shows that the energy density saturates quickly with voltage in the lattice model. In the off-lattice model/3D simulations, $E$ increases much slower even at high applied potential differences due to capacitance’s slower (power-law) decay. Plugging \eq{eq:C:asymp} into \eq{eq:E} shows that the stored energy diverges logarithmically, \ie, it increases illimitably with increasing the voltage. 

Our calculations also show that the energy storage is enhanced by increasing the pore ionophobicity, although the capacitance peaks show the opposite trend (\fig{fig:IL:charging}, middle and bottom rows). More ionophobic pores extend or shift the region of active charging to higher voltages, leading to higher stored energies, as pointed out elsewhere \cite{kondrat:nh:16}.

\subsection{Ion-size asymmetry}
\label{sec:res:ILasym}

\begin{figure}[t]
\begin{center}
    \includegraphics[width=0.85\textwidth]{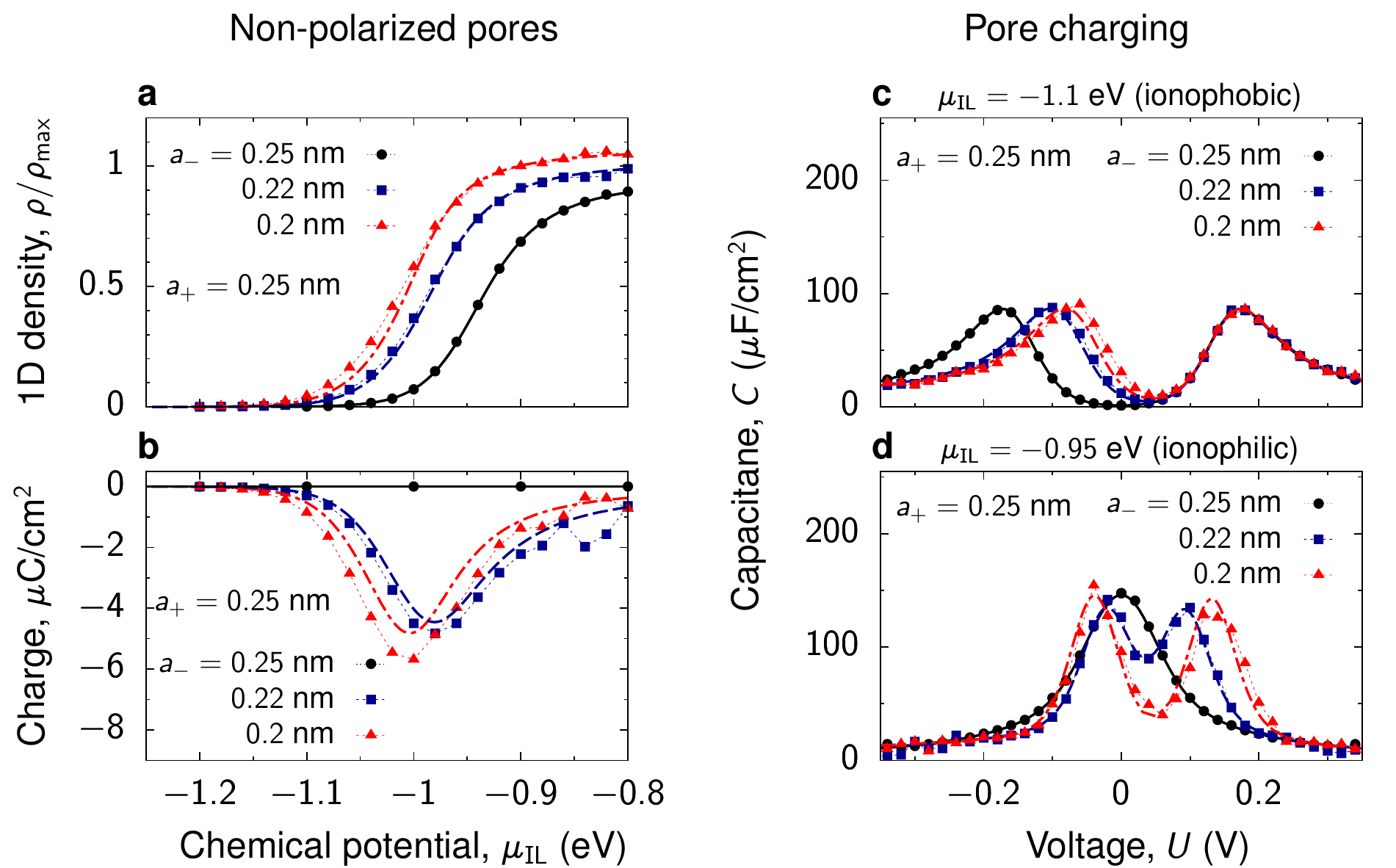}
	\caption{ \ftitle{Ion-size asymmetry.} 1D density \fsub{a} and accumulated charge \fsub{b} as functions of chemical potential $\muIL$ for a few values of the anion radius $\ionr_-$ and cation radius $\ionr_+=\SI{0.25}{\nano\meter}$. \fsub{c,d} Capacitance as a function of applied potential difference for \fsub{c} an ionophobic pore with $\muIL = \SI{-1.1}{\electronvolt}$ and \fsub{d} for an ionophilic pore with $\muIL = \SI{-0.95}{\electronvolt}$. Pore radius $\porer=\SI{0.26}{\nano\meter}$ ($\porer/\ionr_+ = 1.04$), temperature $T=\SI{300}{\kelvin}$, and the in-pore relative dielectric constant $\varepsilon = 2.5$. For ion densities during charging, see \sfig{fig:IL:asym:dens} 
    \label{fig:IL:asym}
}
\end{center}
\end{figure}

A typical ionic liquid has cations and anions of different sizes \cite{Hallett2011Room}. To study how such size asymmetry affects the charging behaviour and to test our 1D off-lattice model, we have considered a few systems with different cation-to-anion size ratios. We found that reducing the anion size increases the overall ion density (\fig{fig:IL:asym}a), which is likely due to the reduced entropic costs of ion packing for smaller anions. \Fig{fig:IL:asym}a also shows that the off-lattice model agrees well with 3D MC simulations. However, the agreement becomes less accurate for decreasing anion size due to the increasing role of the latteral dimensions (across the nanotube), which are accounted for in the 1D model only approximately (\eq{eq:psi_bar}). The deviations are even more pronounced for the accumulated charge at zero voltage, which is non-zero due to ion-size asymmetry (\fig{fig:IL:asym}b).

Interestingly, the magnitude of this zero-potential charge (ZPC) has a peak for moderately filled pores and decreases to zero for more ionophilic and more ionophobic pores. In the limit of strong ionophobicity, the ZPC is zero because the pore is empty in this case. At high ion densities, the size asymmetry becomes entropically less important due to tight packing, and hence the ZPC decreases with increasing the ionophilicity.

Despite the differences in the ZPC, the capacitance as a function of voltage agrees remarkably well within both approaches (\fig{fig:IL:asym}b,c). Clearly, for ionophobic pores, the capacitance at positive polarisations does not depend on the anion size. At negative polarisations, the capacitance peak shifts towards lower voltages for smaller anions due to lowering the entropic costs of charging (pore filling). For ionophilic pores, introducing the ion-size asymmetry transforms the capacitance from bell-shaped to asymmetric camel-shaped (\fig{fig:IL:asym}d). For the increasing ion density due to ion size asymmetry (\fig{fig:IL:asym}a), the minimum appears instead of the maximum because the charging at low voltages becomes driven by co-ion desorption (\sfig{fig:IL:asym:dens}, \cf \fig{fig:IL:charging}c).

\subsection{Effect of non-polar solvent}
\label{sec:res:ILsolv}

We finally discuss charging with mixtures of ions and non-polar solvents. We focus on moderately filled and ionophilic pores, shown in \fig{fig:IL:charging}b,c, and discuss how the charging behaviour changes upon the addition of solvent. For simplicity, in all calculations, we have taken solvent molecules of the same size as the ions and restricted the ion-solvent interactions to hardcore exclusion. 

\begin{figure}[t]
\begin{center}
    \includegraphics[width=0.8\textwidth]{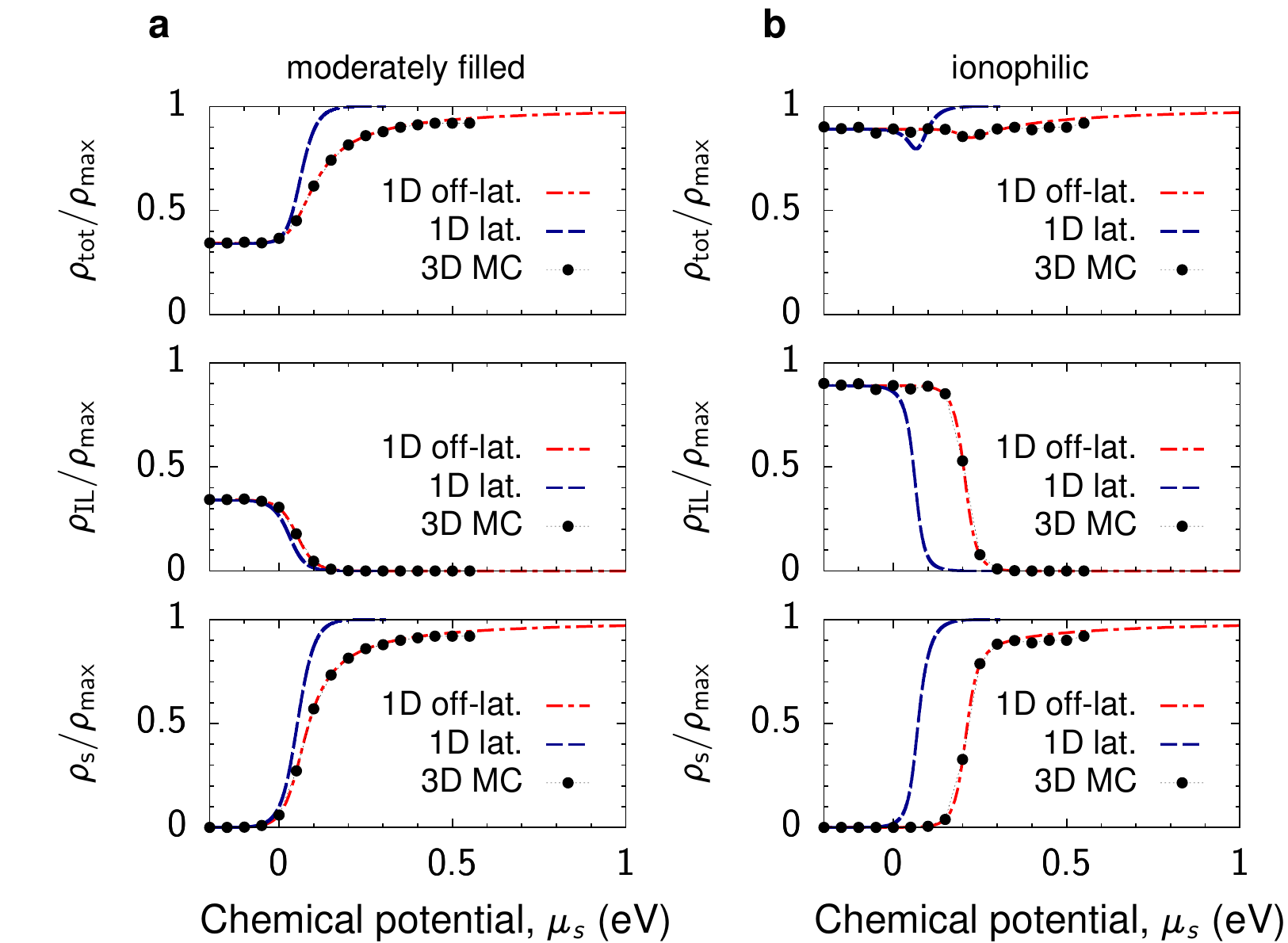}
	\caption{\ftitle{Ionic liquid-solvent mixtures in uncharged metallic nanotubes.} 1D ion densities are shown as functions of the solvent chemical potential $\mus$. Ion diameter $\ionr=\SI{0.25}{\nano\meter}$ and pore radius $R=\SI{0.26}{\nano\meter}$ ($R/a = 1.04$), temperature $T=\SI{300}{\kelvin}$ and dielectric constant $\varepsilon = 2.5$. The chemical potentials of ions are $\muIL=\SI{-0.95}{\eV}$ in \fsub{a} and $\muIL=\SI{-0.8}{\eV}$ in \fsub{b}, giving the ion densities $\rhoIL = 0.34\rhomax$ and $0.89\rhomax$, respectively, in the absence of solvent. The corresponding chemical potential in the lattice model are $\muIL=\SI{0.9608}{eV}$ and $\mus = \SI{0.0261}{eV}$, respectively. Compare with \fig{fig:IL:charging}. The results for the ionophobic pore are shown in \sfig{fig:solv:phobic}.
    \label{fig:solv:nonpol}
}
\end{center}
\end{figure}

\Fig{fig:solv:nonpol} shows various densities as functions of the solvent chemical potential $\mus$. Clearly, increasing $\mus$ expels ions from a pore, even if it is strongly ionophilic. Again, there is an excellent agreement between the 1D off-lattice model and 3D MC simulations, but there are significant differences with the lattice model, particularly for ionophilic pores.

\begin{figure}[th]
\begin{center}
    \includegraphics[width=0.9\textwidth]{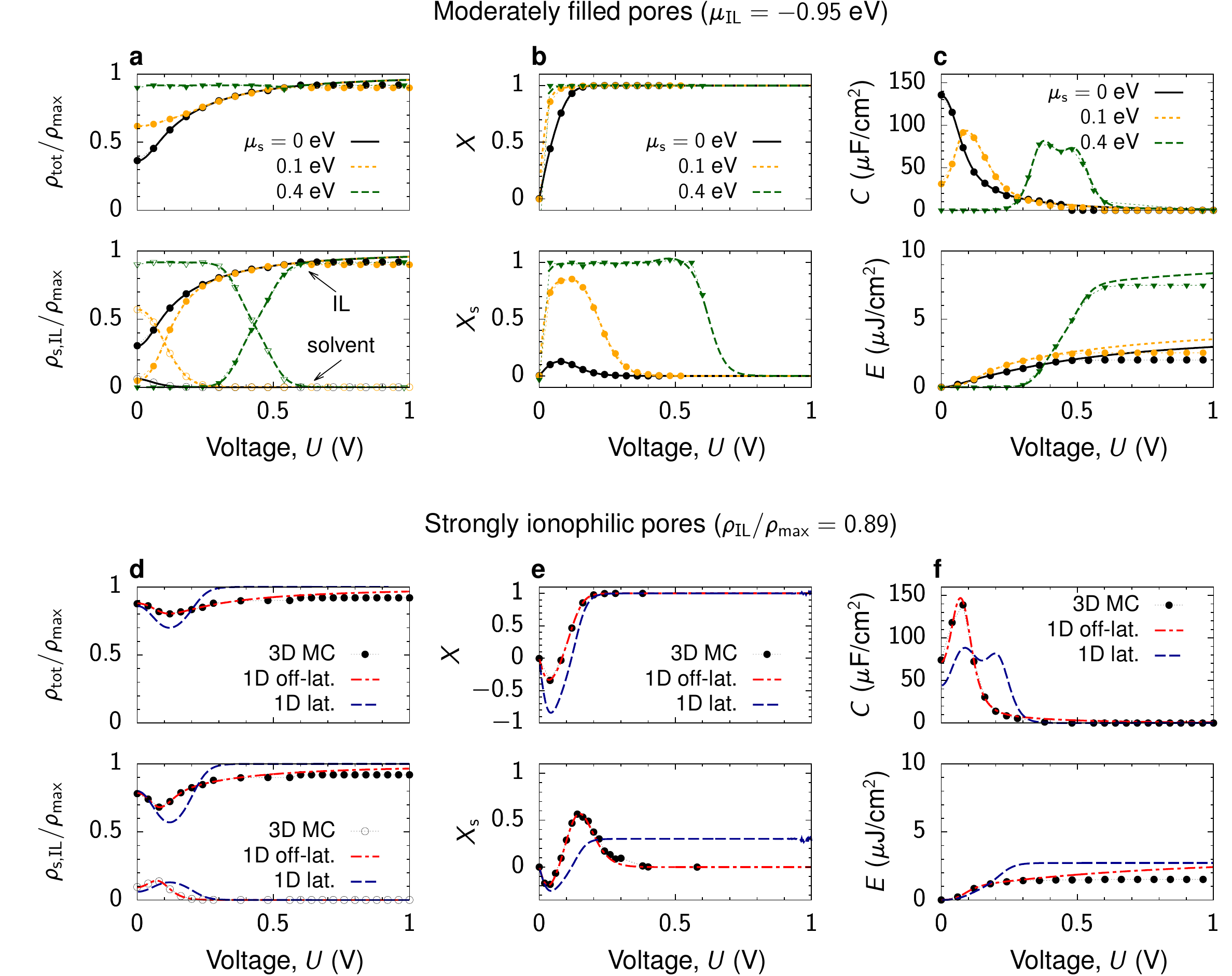}
	\caption{\ftitle{Charging nanotubes with ionic liquid-solvent mixtures.} \fsub{a-c} Charging a nanotube moderately filled with ions in the absence of solvent. Symbols show the results of 3D MC simulations and the lines correspond to the results of the off-lattice model. \fsub{a} Total ion density (upper panel) and solvent and ion densities (lower panel), \fsub{b} charging parameters $\XIL$ and $\Xs$ (\eqs{eq:X}{eq:Xs}), and \fsub{c} capacitance (upper panel) and stored energy density (lower panel) as functions of voltage for a few values of the solvent chemical potential $\mus$. \fsub{d-f} Charging the strongly ionophilic pore (\fig{fig:IL:charging}c and \fig{fig:solv:nonpol}b) from 3D MC simulations and 1D off-lattice and lattice models. \fsub{d}~Total density (upper panel) and solvent and ion densities (lower panel), \fsub{e} charging parameters, and \fsub{f} capacitance and stored energy density as functions of voltage. In all plots, the ion radius $\ionr=\SI{0.25}{\nano\meter}$ and pore radius $R=\SI{0.26}{\nano\meter}$ ($R/a = 1.04$), temperature $T=\SI{300}{\kelvin}$ and dielectric constant $\varepsilon = 2.5$. In the 3D MC simulations and off-lattice model, the IL chemical potential of ions $\muIL=\SI{-0.95}{\eV}$ in  \fsub{a-c} and $\muIL=\SI{-0.8}{\eV}$ in  \fsub{d-f}, giving the IL densities $\rhoIL = 0.34\rhomax$ and $0.89\rhomax$, respectively, in the absence of solvent ($\mus \to -\infty$). In \fsub{d-f}, the solvent chemical potential $\mus = \SI{0.17}{eV}$. In the lattice model (panels \fsub{d-f}), we chose the chemical potentials to provide the same zero-voltage ion and solvent densities as in 3D MC simulations; their values are $\muIL=\SI{0.9608}{eV}$ and $\mus = \SI{0.0261}{eV}$. The results of the lattice model for the moderately filled pore are shown in \sfig{fig:solv:charging:lat}.
    \label{fig:solv:charging}
	}
\end{center}
\end{figure}

\Fig{fig:solv:charging}a shows the densities as functions of voltage for a few values of $\mus$, corresponding to different degrees of pore solvophilicity. This figure demonstrates that the charging proceeds via adsorption of ions and expulsion of solvent. To characterise charging mechanisms quantitatively, we have calculated the charging parameter \cite{forse:jacs:16:chmec, breitsprecher:jcp:17:mcmd, Groda2021Superionic} (see \eq{eq:cap} for the definition of $C(u)$)
\begin{align}
	\label{eq:X}
	X(u) = \frac{e}{2 \pi \porer C(u)}\frac{d\rho_\IL}{du} 
	= \frac{\beta e^2}{2\pi \porer C(u) L }\left[ 
		\langle N_+^2 - N_-^2 \rangle - (\bar N_+^2 - \bar N_-^2) 
		\right].
\end{align}
This parameter is zero when the charging is due to a one-to-one swapping of co-ions for counter-ions, while $\XIL=1$ ($\XIL=-1$) for pure counter-ion adsorption (co-ion desorption). 

To elucidate the role of solvent in the charging process, we introduce the second charging parameter (note the minus sign)
\begin{align}
	\label{eq:Xs}
	X_\solv(u) = - \frac{e}{2\pi \porer C(u)}\frac{d\rho_\solv}{du} 
	= - \frac{\beta e^2}{2\pi \porer C (u) L}
	\left[ \langle N_\solv (N_+ - N_-) \rangle - \bar N_\solv(\bar N_+ - \bar N_-) \right],
\end{align}
where $\bar N_\solv = \langle N_\solv\rangle$ is the average number of solvent molecules. Parameter $\Xs$ quantifies how much of the charge accumulation is due to the desorption or adsorption of solvent molecules. For instance, $\Xs=1$ and $\XIL = 1$ means that adsorption of counter-ions proceeds via expulsion of solvent, that is, the actual charging mechanism is solvent--counter-ion swapping. 

Charging parameters \eqref{eq:X} and \eqref{eq:Xs} are plotted in \fig{fig:solv:charging}b. Parameter $\XIL$ shows that in all cases (at non-zero voltages), the charging is driven by adsorption of counter-ions ($X\approx 1$), while $\Xs > 0$ means that the desorption of solvent molecules in part assists it. This behaviour is most pronounced for the most solvophilic pore considered ($\mus=\SI{0.4}{eV}$ in \fig{fig:solv:charging}b). In this case, $\Xs \approx 1$ in the whole voltage range, except for large voltages, at which the pore becomes entirely free of solvent. Clearly, the contribution from ion-solvent swapping decreases for more solvophobic pores and $\Xs$ becomes significantly smaller unity ($\mus=\SI{0}{eV}$ and $\SI{0.1}{eV}$ in \fig{fig:solv:charging}b).

\Fig{fig:solv:charging}c shows the capacitance and energy density stored in the nanopores. The capacitance experiences a transformation from bell-shaped to camel-shaped as the solvent density increases. At very high solvent concentrations, the pore becomes effectively ionophobic, and the camel-shaped capacitance transforms into a capacitance with four peaks (two at positive and two at negative voltages). The first peak appears when the system overcomes an energy barrier to expel solvent. The second peak develops when the pore becomes free of solvent, and the charging becomes due to overcoming the repulsion between the counter-ions. For the most solvophobic pore, the stored energy is a few-fold higher compared to the nanopores with low or moderate solvent occupancies (\fig{fig:solv:charging}c, bottom plot). This is so because solvophobic pores effectively shift the charging to higher voltages \cite{rochester:jpcc:16}, enhancing the energy storage \cite{kondrat:nh:16}.

Charging the strongly ionophilic pore proceeds by expelling co-ions (\fig{fig:solv:charging}d). Perhaps surprisingly, this process is accompanied by the adsorption of solvent, so that charging parameter $\Xs$ becomes negative (\fig{fig:solv:charging}e). Note that $\Xs > -1$, which means that the solvent adsorption assists only part of the co-ion desorption. Although the lattice model captures this behaviour qualitatively, there are significant quantitative differences. In particular, the charging parameter $\Xs$ saturates at a non-zero value, unlike in the simulations and off-lattice model. The capacitance in the lattice model shows two peaks instead of one (considering only positive voltages), and the stored energy density saturates much more quickly with voltage than in the off-lattice model and 3D MC simulations (\fig{fig:solv:charging}f).

\section{Conclusions}

We have developed an exactly-solvable one-dimensional off-lattice model to study charge storage in conducting single-file pores. We applied this model to neat ionic liquids, including equally-sized and size-asymmetric ions, and to mixtures of ions and non-polar solvents. The model shows a remarkably good quantitative agreement with three-dimensional Monte Carlo simulations for pore radii smaller than $\approx 1.4 \ionr$ ($\ionr$ is the ion radius) and captures the correct behaviour qualitatively also for wider pores. We compared these results with the results of the corresponding lattice model and found that it describes the nanotube charging appropriately only for ionophobic pores at low voltages. This difference is due to entropic effects, which the lattice model does not capture adequately. The off-lattice model is versatile and can be applied to other systems, \eg to include (short-ranged) van der Waal interactions or mixtures of ionic liquids. \label{ref2:p6}One can potentially extend this model to wider pores (\eg, pores accommodating two or three rows of ions, similarly to the lattice model of \myref{ZaboronskyKornyshev2020Ising}) or to polar solvents. Although these models may not be analytically tractable (see, \eg, \myref{fantoni2017one}), they may nevertheless be used for computationally inexpensive numerical analyses.

With analytical calculations and simulations, we found that the capacitance can be bell-shaped (one peak at the potential of zero charge), camel-shaped (two peaks), or have four peaks. We did not observe a bird-shaped (three-peak) capacitance reported in the literature \cite{Alam2008, Cruz2018, Cruz2021Capillary, Cruz2021Capacitance}, which might be because we considered solvent molecules and ions of equal sizes \cite{rochester:jpcc:16} or due to the absence of van der Waals interactions \cite{Cruz2018, Cruz2021Capillary, Cruz2021Capacitance}. The camel and bell-shaped capacitances emerged at low and high ion densities, respectively, similarly to ionic liquids at flat electrodes \cite{dicaprio:03, dicaprio:04, kornyshev:07, Alam2007a, klos:jpcc:10, Henderson2013tail}. Counter-intuitively, at very high densities, the capacitance shape became once again camel-shaped (\fig{fig:IL:charging}), which we associated with co-ion desorption, not occurring at flat electrodes. We observed a four-peak capacitance only for highly solvophilic electrodes (\fig{fig:solv:charging}). However, the lattice model predicted the four-peak capacitance in a much wider range of parameters, in contrast to the off-lattice model and simulations. Transformations between various capacitance shapes could be induced by changing pore ionophilicity (\fig{fig:IL:charging}), ion-size asymmetry (\fig{fig:IL:asym}), and pore solvophilicity (\fig{fig:solv:charging}). \label{ref2:p7}We note, however, that a distribution of pore sizes in real nanoporous electrodes can smear out the capacitance peaks \cite{kondrat:ees:12, kornyshev:fd:14, kondrat:nh:16}, making such transformations challenging to observe experimentally. 

We also found that increasing the ionophobicity or solvophilicity of pores could drastically enhance energy storage (\figs{fig:IL:charging}{fig:solv:charging}), in accord with earlier work \cite{kondrat:nh:16} and  follow-up theoretical and experimental investigations \cite{lian_wu:jpcm:16:Ionophobic, He2020Observation, Gan2021Ionophobic, yin2021carbon}.

Using the off-lattice model, we derived a large-voltage asymptotic expression for the capacitance (\eq{eq:C:asymp}). We found that the capacitance decays as the inverse square of the voltage, \ie, $C\sim u^{-2}$. The lattice model does not capture this behaviour but predicts an exponential decay. Interestingly, the inverse-square law follows entirely from hard-core interactions and does not depend on the details of other interactions between the confined ions. Assuming that this is so also for slit pores and using, in \eq{eq:phen_free_energy}, the scaled-particle results for the entropy of a hard-disk fluid \cite{Helfand1961Theory, Holovko2010Analytical} (instead of \eq{eq:s_1d_rods}), one finds $C \sim u^{-3/2}$. We recall that for flat electrodes, the capacitance decays as $C \sim u^{-1/2}$ \cite{kornyshev:07}, \ie, it shows the slowest decay among the three geometries, as one may expect. Thus, the large-voltage dependence of capacitance may indicate pore shapes in complex networks of supercapacitor electrodes. We note, however, that the voltages at which these asymptotic behaviours dominate might be too high for typical ionic liquids (\sfig{fig:IL:charging:asymp}). Nevertheless, it will be exciting to investigate these asymptotic behaviours and their possible crossovers experimentally and with simulation.

\section*{Supplementary Material}

Derivations and exact results for the 1D off-lattice and lattice models. Supplementary plots.

\section*{Data Availability Statement}

The data that support the findings of this study are available from the corresponding author upon reasonable request.

\bibliography{supercaps,exact}

\end{document}

%% file: defines.tex
\def\papertitle{Capacitive energy storage in single-file pores: Exactly-solvable models and simulations}

\newlength{\figwidth}

\newif\ifOnecolumn
  \ifx\Onecolumn\undefined
  \Onecolumntrue
  
\else
  \Onecolumnfalse
\fi

\newif\ifTwocolumn
  \ifOnecolumn
   \Twocolumnfalse
\else
  \Twocolumntrue
\fi

\usepackage[draft,inline,nomargin]{fixme}

\fxsetup{theme=color,mode=multiuser}

\FXRegisterAuthor{sk}{ask}{\color{red}SK}
\FXRegisterAuthor{tv}{atv}{\color{blue}TV}

\usepackage{xspace}
\newcommand{\latin}[1]{{\it #1}}
\newcommand{\ie}{\latin{i.e.}\@\xspace}
\newcommand{\eg}{\latin{e.g.}\@\xspace}
\newcommand{\cf}{\latin{cf.}\@\xspace}
\newcommand{\etc}{\latin{etc.}\@\xspace}

\newcommand{\fig}[1]{Fig.~\ref{#1}}
\newcommand{\figs}[2]{Figs.~\ref{#1} and \ref{#2}}
\newcommand{\Fig}[1]{Figure~\ref{#1}}

\newcommand{\sfig}[1]{Fig.~\ref{sm:#1}}
\newcommand{\sfigs}[2]{Figs.~\ref{sm:#1} and \ref{sm:#2}}

\newcommand{\sect}[1]{Sec.~\ref{#1}}
\newcommand{\sects}[2]{Sec.~\ref{#1} and \ref{#2}}

\newcommand{\ssect}[1]{Section~\ref{sm:#1} in the Supplementary Material}

\newcommand{\ftitle}[1]{{\bf #1}}
\newcommand{\fsub}[1]{({\bf #1})}

\newcommand{\myref}[1]{ref.~\cite{#1}}
\newcommand{\myrefs}[1]{refs.~\cite{#1}}

\newcommand{\eq}[1]{eq.~\eqref{#1}}
\newcommand{\eqs}[2]{eqs.~\eqref{#1} and \eqref{#2}}
\newcommand{\eqss}[1]{eqs.~\eqref{#1}}
\newcommand{\Eq}[1]{Equation~\eqref{#1}}
\newcommand{\Eqs}[2]{Equations~\eqref{#1} and \eqref{#2}}

\newcommand{\rhomax}{\rho_{\mathrm{max}}}
\newcommand{\solv}{\mathrm{s}}
\newcommand{\IL}{\mathrm{IL}}

\newcommand{\rhos}{\rho_{\solv}}
\newcommand{\rhoIL}{\rho_{\IL}}

\newcommand{\rhoc}{\rho_{\mathrm{c}}}

\newcommand{\mus}{\mu_{\solv}}
\newcommand{\muIL}{\mu_{\IL}}

\newcommand{\Xs}{X_{\solv}}
\newcommand{\XIL}{X}

\newcommand{\porer}{R}
\newcommand{\ionr}{a}
\newcommand{\solvr}{a_\solv}

\newcommand{\rr}{\boldsymbol{r}}
\newcommand{\e}{\mathrm{e}}
\newcommand{\lat}{\mathrm{lat}}

\newcommand{\oneD}{\mathrm{1D}}

\newcommand{\Tr}{\mathrm{Tr}}

\newcommand{\XiL}{\widetilde\Xi_\oneD}
\newcommand{\Ncells}{\mathcal{N}}

\newcommand{\lB}{\lambda_B}

